\documentclass[twocolumn,showpacs,preprintnumbers,amsmath,amssymb]{revtex4}
\usepackage{dcolumn}
\usepackage{bm}
\usepackage{graphicx}

\def\w0{\omega_0}
\def\k0{k_0}
\def\bk{{\bf k}}
\def\bkp{{\bf k}'}
\def\ad{a^\dagger}
\def\bkj{b_{\bk j}}
\def\bkjd{b_{\bk j}^\dagger}
\def\bkjp{b_{\bk ' j'}}

\def\fkj{f_{\bk j}}
\def\fkjp{f_{\bk 'j'}}
\def\Bkj{B_{\bk j}}
\def\Bkjd{B_{\bk j}^\dagger}
\def\Bkjp{B_{\bk 'j'}}
\def\tkj{t_{\bk j}}
\def\rkj{r_{\bk j}}
\def\Tkjpkj{T_{\bk j}^{\bk 'j'}}
\def\Rkjpkj{R_{\bk j}^{\bk 'j'}}
\def\lkj{\lambda^{\bk j}}
\def\takj{\tau^{\bk j}}
\def\mkjpkj{\mu_{\bk j}^{\bk 'j'}}
\def\ekjpkj{\eta_{\bk j}^{\bk 'j'}}
\def\ekj{\hat{e}_{\bk j}}

\def\bkp{{\bf k}'}
\def\br{{\bf r}}

\begin{document}

\title{Harmonic oscillator model for the atom-surface Casimir-Polder interaction energy}

\author{Roberto Passante, Lucia Rizzuto, Salvatore Spagnolo}
\affiliation{Dipartimento di Fisica dell'Universit\`{a} degli Studi di Palermo and CNISM, Via Archirafi 36, I-90123 Palermo, Italy}

\email{roberto.passante@unipa.it}

\author{Satoshi Tanaka}
\affiliation{Department of Physical Science, Osaka Prefecture University, Sakai 599-8531, Japan}

\author{Tomio Y. Petrosky}
\affiliation{Center for Complex Quantum Systems, The University of Texas at Austin, Austin, Texas 78712, USA}


\begin{abstract}
In this paper we consider a quantum harmonic oscillator  interacting with the electromagnetic radiation field in the presence of a boundary condition preserving the continuous spectrum of the field, such as an infinite perfectly conducting plate. Using an appropriate Bogoliubov-type transformation we can diagonalize exactly the Hamiltonian of our system in the continuum limit and obtain non-perturbative expressions for its ground-state energy. From the expressions found, the atom-wall Casimir-Polder interaction energy can be obtained, and well-know lowest-order results are recovered as a limiting case. Use and advantage of this method for dealing with other systems where perturbation theory cannot be used is also discussed.
\end{abstract}

\maketitle

\section{\label{sec:1}Introduction}

Many physical phenomena, including Casimir and Casimir-Polder forces or the Lamb shift, are related to the energy shift of the ground state of a system with a discrete spectrum interacting with a field with a continuum spectrum \cite{CTDRG92}. Casimir and Casimir-Polder interactions have recently received much attention in the literature being a direct manifestation of the quantum properties of the electromagnetic field, in particular of the existence of vacuum fluctuations, even at a macroscopic level \cite{Milonni07}.  They also have raised great interest for their  importance in applications to nanotechnological devices such as micro-electromechanical systems (MEMS) \cite{CMC11}.

The Casimir-Polder atom-surface potential is a long-range interaction between a neutral atom or molecule (or in general a polarizable body) and a conducting wall in vacuo \cite{CP48}, and it is usually considered for systems in their ground state.
This potential arises from the interaction of the atom with the vacuum fluctuations of the quantum electromagnetic field modified by the presence of the conducting wall: this yields an interaction energy depending on the atom-wall distance and thus a force on the atom.
This force has a pure quantum origin, and for ground state atoms can be obtained by a second-order perturbative calculation \cite{Milonni94}.

A similar force exists also for atoms in an excited state, with a different dependence on the atom-wall distance due to the presence of atom-field resonances \cite{Barton87,MPRSV08}.
Perturbative calculations of Casimir forces for excited states may encounter difficulties due to resonance divergences. There are indeed some controversies in the literature when Casimir-Polder forces are considered for systems in their excited state, in particular when a resonant energy transfer between two identical atoms can occur  \cite{BLMN03,HSH11}.
Also perturbative calculations of atomic radiative energy shifts encounter difficulties related to vanishing energy denominators  \cite{PP08}.
Time dependence of the atom-wall Casimir-Polder interaction leading to a dynamical Casimir-Polder effect has been also recently considered in the literature \cite{VP08,MVP10}.

Nonperturbative calculations may avoid the difficulties arising for diverging energy denominators and, at the same time, be useful for obtaining Casimir-like interactions for other physical systems where the coupling with the field could be large or when the density of photon states is large.
For example, a large photon density of states occurs  near the band-edge of a photonic crystal \cite{John}, and this may require nonperturbative approaches.
In condensed matter physics, the density of states of one-dimensional systems such as a quantum wire has the well-known Van Hove singularity near the edges of the band, and its role in enhancing decay rates \cite{PTG05} or charge transfer from one or multiple impurities to the wire \cite{TGP06,TGOP07} have been recently investigated with nonperturbative methods. An analogous effect investigated in the last years is the increase of the Lamb shift of an excited level of an atom in a photonic crystal \cite{WKG04}. Non-perturbative methods with a dressed-state approach have been used for describing the radiative spontaneous decay from an excited state \cite{FHM02,FHM05} or the evolution of an unstable system \cite{Longhi09}.

In a previous paper we have used a method based on Bogoliubov-like transformations in order to obtain the Casimir-Polder interaction between two atoms, where the atoms were modeled as three-dimensional harmonic oscillators \cite{CKP05}.
This method based on Bogoliubov-type transformation was originally introduced in Ref. \cite{AGPP98}  for a one-dimensional scalar field and applied to study the dynamics of an oscillator interacting with a scalar field \cite{KPPP00}.
In this paper we shall develop a similar method for treating the atom-wall Casimir-Polder interaction approximating the atom as a quantum harmonic oscillator and diagonalizing the interacting Hamiltonian of the system.
We obtain two exact expressions (within dipole approximation) for the ground-state energy, from which the oscillator-wall Casimir-Polder energy can be obtained.
One of the two exact energy shifts obtained is in terms of the resolvent, and the other in terms of the function resulting from the factorization problem of the resolvent. The known second-order expression of the force is recovered as a limiting case from both expressions.
Also, we stress that the validity of our method is not limited to this case and can be applied to any boundary condition that yields a continuous spectrum of the field. In fact, the boundary condition enters only in the specific form of the mode functions of the field, and consequently in oscillator-field coupling constant: our diagonalization procedure of the Hamiltonian is valid for a general expression of the coupling constant. We specify its form only at the very end of the calculation. Moreover, having obtained a diagonal form of the Hamiltonian, could make easier dealing with different states of the interacting system, such as thermal states, excited states or in general non-equilibrium states.

Examples of possible applications of our method to other physical systems, such as impurities in low-dimensional systems in condensed matter physics or dispersion interactions between atoms or quantum dots in a photonic crystal, are also briefly discussed.

The outline of this paper is as follows. In Section \ref{sec:2} we introduce our Hamiltonian model and the improper Bogoliubov-type transformation that diagonalizes the Hamiltonian of the oscillator interacting with the radiation field inclusive of the boundary condition; we then obtain two alternative (but equivalent) exact expressions for the ground-state energy shift.
In Section \ref{sec:3} we apply the results of the previous Section in order to obtain the oscillator-surface Casimir-Polder potential energy, recovering the known lowest-order result as a limiting case; we also show the importance of the resonance poles in the expressions obtained for the energy shift of the ground state. In Section \ref{sec:4} we summarize our results and discuss their possible relevance in different physical systems where strong coupling constants or high density of states require the use of nonperturbative methods.

\section{\label{sec:2}The Hamiltonian model and the Bogoliubov transformation}

We consider a quantum harmonic oscillator with frequency $\w0 = c\k0$, described by annihilation and creation operators
$a$ and $\ad$, interacting with the quantum electromagnetic field described by the annihilation and creation operators $\bkj$ and $\bkjd$. The presence of the boundary conditions is mathematically included in the form of the coupling constant $\fkj$ and of the field mode functions $\mathbf{\tilde{f}}_{\bk j} (\br)$.

The Hamiltonian of our system is
\begin{eqnarray}
H &=& \hbar c \k0 \ad a + \sum_{\bk j} \hbar c k \bkjd \bkj  \nonumber\\
&&
+ \sum_{\bk j} \fkj(\br_A) \left( a + \ad \right) \left(  \bkj - \bkjd \right) \;,
\label{eq:1}
\end{eqnarray}
where the third term is the atom-field interaction in the multipolar coupling scheme within dipole approximation. Phases have been chosen such that the coupling constants $\fkj$, which are evaluated at the atom/oscillator's position $\br_A$, be pure imaginary; they are given by
\begin{equation}
\fkj (\br)= i\sqrt{\frac{2\pi \hbar ck}V}\mbox{\boldmath $\mu$} \cdot \mathbf{\tilde{f}}_{\bk j} (\br) \;,
\label{eq:2a}
\end{equation}
where $\mbox{\boldmath $\mu$}$ is the matrix element (real) of the oscillator dipole moment and $V$ the quantization volume.

In (\ref{eq:2a}), $\mathbf{\tilde{f}}_{\bk j} (\br)$ are the mode functions of the electromagnetic field and take into account possible boundaries. If a conducting plate at $z=0$ is present, they can be obtained from those of a perfectly conducting cubical cavity of volume $V=L^3$ with walls  ($-L/2<x,\; y<L/2$, \;$0<z<L$)
\begin{widetext}
\begin{equation}\label{eq:2b}\begin{split}
&(\mathbf{\tilde{f}}_{\mathbf{k}j})_x=\sqrt{8}(\hat{e}_{\mathbf{k}j})_x
\cos\biggl[k_{x}\biggl(x+\frac L2\biggl)\biggl]
\sin\biggl[k_{y}\biggl(y+\frac L2\biggl)\biggl]\sin\bigl(k_{z}z\bigl) \;,\\
&(\mathbf{\tilde{f}}_{\mathbf{k}j})_y=\sqrt{8}(\hat{e}_{\mathbf{k}j})_y
\sin\biggl[k_{x}\biggl(x+\frac L2\biggl)\biggl]
\cos\biggl[k_{y}\biggl(y+\frac L2\biggl)\biggl]\sin\bigl(k_{z}z\bigl)\;, \\
&(\mathbf{\tilde{f}}_{\mathbf{k}j})_z=\sqrt{8}(\hat{e}_{\mathbf{k}j})_z
\sin\biggl[k_{x}\biggl(x+\frac L2\biggl)\biggl]
\sin\biggl[k_{y}\biggl(y+\frac L2\biggl)\biggl]\cos\bigl(k_{z}z\bigl) \;,
\end{split}\end{equation}
\end{widetext}
where $k_x=l\pi/L$,\; $k_y=m\pi/L$,\; $k_z=n\pi/L$ ($l,m,n=0,1,2,\dots$) and $\ekj$ are polarization unit vectors \cite{PT82}.
In order to switch from the cavity to the wall at $z=0$, at the end of
the calculations one has to take the limit $L\rightarrow\infty$.

Our Hamiltonian \eqref{eq:1} also includes counter-rotating terms yielding virtual processes. It is well known that they are essential for correctly obtaining radiative processes such as the Lamb shift or Casimir forces for ground-state systems \cite{CPP95}.
This is the great advantage of modeling the atom or polarizable body as a harmonic oscillator: it allows an exact diagonalization of the total Hamiltonian, even maintaining counter-rotating terms. Although at the end we shall focus our discussion on the specific case of the atom-surface Casimir-Polder force, our method is more general and can be applied to any boundary conditions, provided the field spectrum is continuous. In fact, the specific boundary condition enters only in the form of $\fkj$ in \eqref{eq:1}, and we shall specify it only at the very end of our calculations.

Following a similar method to that used in \cite{KPPP00,CKP05}, we introduce {\it new} bosonic operators related to the {\it old} ones by the following Bogoliubov-like transformations
\begin{eqnarray}
\Bkj &=& \tkj^* a + \rkj^* a^\dagger + \sum_{\bkp j'} \left( {\Tkjpkj}^* \bkjp +{\Rkjpkj}^* \bkjp^\dagger \right) ,\nonumber  \\
\Bkj^\dagger &=& \tkj a^\dagger + \rkj a + \sum_{\bkp j'} \left( \Tkjpkj \bkjp^\dagger +\Rkjpkj \bkjp \right) .
\label{eq:3}
\end{eqnarray}
We determine the coefficients in \eqref{eq:3} by requiring that the transformed operators satisfy free-field commutation relations with the Hamiltonian
\begin{equation}
\left[ H , \Bkj \right] = -\hbar ck \Bkj
\label{eq:4}
\end{equation}
and its Hermitian conjugate relation. In \eqref{eq:4} the frequencies associated to the {\it new} operators are the same as the  {\it old} ones. This is strictly true in the continuum limit ($V=L^3 \to \infty$), and in the discrete case there are corrections $O(1/V)$. Even if for simplicity we keep the notation with discrete sums, it is understood that we shall take the continuum limit at the end of our calculations.

Transformations \eqref{eq:3} are indeed improper Bogoliubov transformations, because there is not a one-to-one correspondence between {\it new} and {\it old} operators: in the diagonalized form, the discrete state disappears in the continuum \cite{AGPP98}. This is similar to what happens also in the Friedrichs model, describing a two-level system coupled with a field continuum in the rotating wave approximation \cite{Friedrichs48,PPT91}.
Use of  \eqref{eq:3} in \eqref{eq:4} yields a set of coupled equations for the coefficients, from which we obtain
\begin{eqnarray}
\tkj &=& -(k+\k0 )G(k) \frac 1{\hbar c}\fkj  \;,\\
\Tkjpkj &=& \delta_{\bk \bkp} \delta_{jj'} + \frac 1{(\hbar c)^2}\frac {2\k0}{k-k'}G(k)\fkj \fkjp \;,\\
\rkj &=& \frac {k-\k0}{k+\k0} \tkj \;,\\
\Rkjpkj &=& \frac {k'-k}{k'+k} \Tkjpkj \;,\nonumber\\
\label{eq:5}
\end{eqnarray}
with the resolvent given by
\begin{equation}
\left( G(k) \right)^{-1} = \k0^2 -k^2 -\frac 1{(\hbar c)^2}\sum_{\bkp j'}
\frac {4\k0 k'}{k^2-k'^2}\fkjp^2 \;.
\label{eq:6}
\end{equation}

If the coupling constant is such that
\begin{equation}
\k0^2 -\frac 1{(\hbar c)^2}\sum_{\bkp j'}\frac {4\k0}{k'}\mid \fkjp \! \mid^2 \;>0
\label{eq:6a}
\end{equation}
the resolvent $G(z)$ is analytic in the complex $z$ plane apart for a cut on the real axis. The resolvent can be analytically continued on the second Riemann sheet, where it can have complex poles related to energy shifts and decay rates \cite{KPPP00,GP11}.

After some algebraic calculations, we can also obtain the inverse relations expressing the {\it old} operators in terms of the {\it new} ones
\begin{eqnarray}
a^\dagger &=& \sum_{\bk j} \left( \lkj \Bkjd + \takj \Bkj \right)
\label{eq:6b} \;, \\
\bkjd &=& \sum_{\bkp j'} \left( \mkjpkj \Bkjp^\dagger + \ekjpkj \Bkjp \right)\;,
\label{eq:6c}
\end{eqnarray}
where the coefficients are given by
\begin{eqnarray}
\lkj &=& (k + \k0 ) G(k) \frac 1{\hbar c} \fkj  \;, \\
\mkjpkj &=& \delta_{\bk \bkp } \delta_{jj'} + \frac {2\k0}{k-k'} G(k) \frac 1{(\hbar c)^2} \fkj \fkjp  \;,\\
\ekjpkj &=& - \frac {2\k0}{k+k'} G^* (k) \frac 1{(\hbar c)^2} \fkj^* \fkjp  \;, \\
\takj &=& -(k - \k0 ) G^* (k) \frac 1{\hbar c} \fkj^* \;.
\label{eq:7}
\end{eqnarray}

Substitution of \eqref{eq:6b}, \eqref{eq:6c} and their Hermitian conjugate with the coefficients above into the Hamiltonian \eqref{eq:1}, after lengthy algebraic calculations, allows to obtain $H$ in a diagonal
form in terms of the new operators plus an energy shift $E_g$  as
\begin{equation}
H = \sum_{\bk j} \hbar c k \Bkjd \Bkj + E_g  \;.
\label{eq:8}
\end{equation}

The energy shift $E_g$ is given by
\begin{eqnarray}
&E_g& = \frac {\k0}{\hbar c} \left[ \sum_{\bk j} (k-\k0)^2 \mid G(k) \mid^2 \mid \fkj \mid^2 \right.
\nonumber \\
&-& \! \! \! \! \left. \frac{4\k0}{(\hbar c)^2} \sum_{\bk j \bkp j'} \mid \fkj \mid^2 \mid \fkjp \mid^2
\mid G(k') \mid^2 \frac {k+2k'}{(k+k')^2} \right] \; .
\label{eq:8a}
\end{eqnarray}

This expression for the ground-state energy shift is exact, and valid at any order in the coupling constant.
As we will see in the next section, when using this expression much attention must be paid because of the singularities of the resolvent $G(z)$ in the complex plane. For example, the second term on the RHS of \eqref{eq:8a}, in the limit of vanishing coupling constant has two second-order poles at $k=\pm \k0$; these second-order poles separate each other when higher-order terms in the coupling are included.

Just for the purpose to obtain $E_g$, there is another way similar to that used in \cite{KPPP00}, main difference in our case being the presence of the boundary condition and having a vector rather than a scalar field. From the general theory of Bogoliubov transformation, the dressed ground state $| 0 \rangle_d$ of the full Hamiltonian $H$ can be obtained by a transformation on the bare ground state (ground state of the unperturbed Hamiltonian $H_0$)  as
\begin{equation}
| 0 \rangle_d = e^V | 0 \rangle
\label{eq:9}
\end{equation}
where $| 0 \rangle$ is the non-interacting ground state and the operator $V$ is a quadratic form of the creation operators. In our case, we have
\begin{eqnarray}
V& =& \left(\frac 1{\hbar c}\right)^2 \sum_{\bk j \bkp j'} k_0 \fkj \fkjp  \eta (k) \eta (k')
\nonumber\\
&&\times \left( \frac 1{2\left( \k0 + \frac {E_g}{\hbar c}\right)} + \frac 1{k+k'} \right)
b_{\bk j}^\dagger b_{\bkp j'}^\dagger \nonumber \\
&&- \frac 1{\hbar c} \sum_{\bk j} \frac {\k0 \fkj \eta (k)}{\k0 +\frac{E_g}{\hbar c}}\bkjd \ad
- \frac {E_g}{2(\hbar c \k0 + E_g )} {\ad}^2 \;,\nonumber\\
\label{eq:10}
\end{eqnarray}
where $\eta (z)$ is the solution of the factorization problem for the resolvent \cite{AGPP98}
\begin{equation}
G(z) = \eta (z) \eta (-z)  \;.
\label{eq:12}
\end{equation}

The function $\eta (z)$ in (\ref{eq:12}) is analytic in the complex $z$ plane apart from a cut in the negative real axis.

The average energy is then obtained by
\begin{equation}
E_g = {\ }_d \! <0|H|0>_d = <0|e^{-V}He^V|0>  \;.
\label{eq:13a}
\end{equation}
After some algebraic calculations, using \eqref{eq:10} we can obtain the expression for the shift of the ground-state energy shift
\begin{equation}
E_g  = \frac {\k0}{\k0 +\frac 1{\hbar c}{E_g}}\frac 1{\hbar c} \sum_{\bk j} \mid \fkj \mid^2 \eta (k)  \;.
\label{eq:14}
\end{equation}

This is an exact equation for the ground-state energy shift $E_g$. We are interested in solutions of \eqref{eq:14} that are
analytic in the coupling constant $\fkj$, that is such that $E_g \to 0$ for $\fkj \to 0$.

We have so obtained two different expressions for the energy shift of the ground state, given by \eqref{eq:8a} and \eqref{eq:14}: the former is in terms of the resolvent $G(z)$ and the latter in terms of the function $\eta (z)$ resulting from the factorization problem of the resolvent. In the next Section we shall use these expressions to recover, at the lowest (second)-order limit, the Casimir-Polder energy between an atom/oscillator and a conducting plate.

\section{\label{sec:3}The ground-state energy and the atom-surface Casimir-Polder potential energy}

We will now show that the expressions of the ground-state energy  \eqref{eq:8a} and \eqref{eq:14} can be used to obtain the Casimir-Polder interaction between an atom, modeled as a harmonic oscillator, and a surface (or any boundary condition preserving the continuous character of the modes of the electromagnetic field). We need to use the appropriate coupling constant $\fkj$. For a perfectly conducting infinite wall, the coupling constant is given by \eqref{eq:2a} with the normal modes in \eqref{eq:2b}. The energy depends on the distance between the oscillator/atom and the conducting wall due to the presence of the functions $\fkj(\br_A)$ evaluated at the atomic position: minus its derivative with respect to $\mid \br_A \mid$ yields the oscillator-wall Casimir-Polder force.
Up to this point this result is exact within the dipole approximation.

We shall now show that the lowest order approximation of $E_g$ will correctly reproduces the well-known second-order atom-wall Casimir-Polder interaction energy.
As mentioned in the previous Section, approximation of \eqref{eq:8a} must be done carefully due to poles of the resolvent $G(z)$ in the second Riemann sheet. We will first show that the term in the second line, which apparently gives a contribution starting from fourth order only, indeed contributes to the second-order energy shift due to the resonance poles of $G(z)$. In fact, when the sum over $\bkp$ is performed in the continuum limit, the poles of $G(z)$ in the second Riemann sheet yield a factor in the denominator proportional to the square of the coupling constant (more specifically to the decay rate of the oscillator), and this makes this contribution a second-order one. We will now show this in the lowest-order approximation.

We have the following relation in the continuum limit between $G^+(k)=G(k+i0)$ and $G^-(k)=G(k-i0)$,  i.e. the extensions of $G(k)$ in the upper and lower complex plane, respectively,
\begin{eqnarray}
&\ & (G^+(k))^{-1} - (G^-(k))^{-1} \nonumber \\
&=& -4\pi i \k0 \frac 1{(\hbar c)^2}\frac V{(2\pi)^3} \sum_j \int \! d\Omega \mid \fkj \mid^2 k^2 \;.
\label{eq:15}
\end{eqnarray}
Using this relation in the sum/integral over $(\bkp j')$ in \eqref{eq:8a}, we obtain
\begin{equation} \label{eq:16} \begin{split}
&\frac 1{(\hbar c)^2}\sum_{\bkp j'} \mid \fkjp \mid^2
\mid G(k') \mid^2 \frac {k+2k'}{(k+k')^2} \\
&= -\frac 1{4\pi i \k0} \int_0^\infty \! \! dk' \frac {k+2k'}{(k+k')^2}
\left( G^-(k') - G^+(k') \right) \\
&\simeq \frac 1{4\k0^2} \frac{k+2\k0}{(k+\k0)^2} \;,
\end{split}
\end{equation}
where we have used (at the lowest order)
\begin{equation} \label{eq:17} \begin{split}
&G^-(k') - G^+(k') \\
&\simeq \frac 1{\k0 +k'} \left( \frac 1{\k0 -k' +i0}- \frac 1{\k0 -k' -i0} \right) \\
&= -2\pi i \frac 1{\k0 + k'} \delta (\k0 -k') \;.
\end{split}
\end{equation}
This explicitly shows that \eqref{eq:17} is indeed of zeroth-order in the interaction due to the resonance poles of the resolvent.
Substitution of \eqref{eq:16} into \eqref{eq:8a} gives then the energy shift approximated to the second order (continuum limit is understood)
\begin{eqnarray}
E_g^{(2)} &\simeq& - \frac 1{\hbar c} \sum_{\bk j} \frac {\mid \fkj (\br_A) \mid^2}{k+\k0} \nonumber\\
&=&-\frac{2\pi}V \sum_{\bk j}\frac k{k+\k0}\mid \mbox{\boldmath $\mu$} \cdot \mathbf{\tilde{f}}_{\bk j} (\br_A) \mid^2  \;,
\label{eq:18}
\end{eqnarray}
that coincides with the results that can be obtained by perturbation theory \cite{Barton87,MPRSV08}.

The same approximated result can be also obtained from the expression \eqref{eq:14} of $E_g$.
By substituting $E_g$ in the RHS of \eqref{eq:14} with its zeroth order value, that is zero, we obtain
\begin{equation}
E_g^{(2)} = \frac 1{\hbar c}\sum_{\bk j} \eta^{(0)} (k) \fkj^2  \;.
\label{eq:19}
\end{equation}
where $\eta^{(0)} (k)$ is the zeroth-order approximation to $\eta (k)$.
Approximating \eqref{eq:6} at the lowest order, we have
\begin{equation}
G(z) \simeq (\k0^2-z^2)^{-1}  \;,
\label{eq:20}
\end{equation}
and taking into account the factorization \eqref{eq:12} of the Green's function and that $\eta(z)$ has a cut in the negative
real axis, we obtain
\begin{equation}
\eta^{(0)} (z) =  -\frac 1{z+\k0} \;.
\label{eq:21}
\end{equation}
Thus the second-order energy shift is
\begin{equation}
E_g^{(2)} \simeq -\frac{2\pi}V \sum_{\bk j}\frac k{k+\k0}\mid \mbox{\boldmath $\mu$} \cdot \mathbf{\tilde{f}}_{\bk j} (\br_A) \mid^2  \;,
\label{eq:22}
\end{equation}
where we have used \eqref{eq:2a} and have explicitly indicated the dependence of $\fkj$ and $\mathbf{\tilde{f}}_{\bk j}$ from the oscillator's position $\br_A$ with respect to the conducting plate. This results coincides with \eqref{eq:18}, of course.

Expression \eqref{eq:18} coincides with the well-known result obtained by perturbation theory at the lowest significant order and in the continuum limit ($V = L^3 \rightarrow \infty$), it yields the known atom-wall interaction energy as $r_A^{-3}$ in the near zone ($r_A \ll \k0^{-1}$) and as $r_A^{-4}$ in the far zone ($r_A \gg \k0^{-1}$)  \cite{Milonni94,MPRSV08}. In a quasi-static approach, where the kinetic energy of nuclei is neglected and translational degrees of freedom are decoupled from the electronic coordinates, the atom-plate Casimir-Polder force can then be obtained as minus the derivative of the energy shift with respect to the oscillator position $F_{CP}=-\partial E_g / \partial d$. This result can be also generalized to more complicated boundary conditions by using the appropriate mode functions in the field operators.

The expression \eqref{eq:8a} of $E_g$ may appear more complicated than \eqref{eq:14}. However, we feel that \eqref{eq:8a} could in general be applied more easily because it involves $G(z)$, for which we have explicit exact and approximated expressions, rather than the function $\eta (z)$.

\section{\label{sec:4}Concluding remarks}

We have considered a quantum harmonic oscillator interacting with the quantum electromagnetic field in the continuum limit and in the presence of boundary conditions. By an improper Bogoliubov-type transformation we have obtained a diagonal form of the Hamiltonian in terms of  {\it new} bosonic operators with an energy shift due to the oscillator-field coupling. Counter-rotating terms are included in the Hamiltonian. Exact expressions for the ground-state energy shift have been obtained using two different methods. As a specific case, we have considered an infinite conducting plate and the resulting Casimir-Polder potential energy between the oscillator and the plate, obtained from the dependence of the ground-state energy shift from the oscillator-plate distance.

We wish to stress that the validity of our method, however, is not restricted to this specific situation, but holds for any boundary conditions yielding a continuous spectrum for the field modes. The particular boundary condition, in fact, enters in the form of the coupling constant $\fkj$ only, and the diagonalization of the Hamiltonian given in Section \ref{sec:2}, as well as the expression of the ground-state energy $E_g$ given in \eqref{eq:8a} and \eqref{eq:14}, are valid for any form of the coupling constant. From the energy shift of the ground state of the interacting system, an exact expression of the Casimir-Polder force acting on the oscillator, due to the presence of the boundary condition, can be obtained. In a quasi-static approach, this immediately yields the Casimir-Polder force on the oscillator. We have also shown that, in the case of a perfectly conducting plate and in the lowest-order approximation in the coupling constant, the well-known expression for the second-order atom-wall Casimir-Polder energy is recovered.

The use of the nonperturbative approach presented in this paper could be particularly relevant in all cases of the interaction of a discrete system with a quantum field, whenever perturbation theory cannot be applied, not necessarily in the framework of quantum optics or quantum electrodynamics. Examples are systems with large coupling constant and/or large density of states for the field modes. An example is the study of radiative processes of atoms or quantum dots in a photonic crystal, when relevant transition frequencies are close to the band edge of the crystal. It is in fact known that in this case the density of states increases and has a divergence at the band-edge, resulting in the necessity of a nonperturbative approach. In the framework of condensed matter physics, a similar situation occurs in the case of impurities in a quantum wire, as a consequence of the van Hove singularity of the electron density of states. We shall discuss applications to these systems in a future publication.

\begin{acknowledgments}
The authors thank R. Messina for interesting discussions on the subject of this paper. The authors acknowledge support from the ESF Research Networking Program CASIMIR. Financial support by Ministero dell'Istruzione, dell'Universit\`{a} e della Ricerca and by Comitato Regionale di Ricerche Nucleari e di Struttura della Materia is also acknowledged.
\end{acknowledgments}

\end{document}